\documentclass[aps,prl,twocolumn,showpacs,superscriptaddress]{revtex4-1}
\usepackage{graphicx}
\usepackage{amsmath}
\usepackage{amssymb}
\usepackage{colordvi}
\usepackage{mathrsfs}
\usepackage{bm}
\usepackage{verbatim}
\usepackage{dcolumn}
\usepackage{bm}
\usepackage{epsfig}
\usepackage{subfigure}
\usepackage{makecell}
\usepackage[colorlinks,allcolors=blue]{hyperref}

\begin{document}
\title{Realization of Kane-Mele Model in $\pmb X\bf{N_4}$-Embedded Graphene ($\pmb X$=Pt, Ir, Rh, Os)}
\author{Haonan Wang}
\affiliation{CAS Key Laboratory of Strongly-Coupled Quantum Matter Physics, and Department of Physics, University of Science and Technology of China, Hefei, Anhui 230026, China}
\affiliation{Department of Physics, Washington University in St. Louis, St. Louis, Missouri 63130, USA}
\author{Qian Niu}%
\affiliation{CAS Key Laboratory of Strongly-Coupled Quantum Matter Physics, and Department of Physics, University of Science and Technology of China, Hefei, Anhui 230026, China}
\author{Zhenhua Qiao}%
\email[Correspondence author:~]{qiao@ustc.edu.cn}
\affiliation{CAS Key Laboratory of Strongly-Coupled Quantum Matter Physics, and Department of Physics, University of Science and Technology of China, Hefei, Anhui 230026, China}
\affiliation{International Center for Quantum Design of Functional Materials, University of Science and Technology of China, Hefei, Anhui 230026, China}
\date{\today}

\begin{abstract}
  Monolayer graphene embedded with transition metal nitride (i.e., $X$N$_4$) has been experimentally synthesized recently, where a transition metal atom together with four nitrogen atoms as a unit are embedded in graphene to form a stable planar single-atom-thick structure. We provide a systematic study on the structural, electronic and topological properties of these $X$N$_4$-embedded graphene by utilizing both first-principles calculations and tight-binding model. We find that $X$N$_4$-embedded graphene ($X$=Pt, Ir, Rh, Os) can open topologically nontrivial band gaps that host \emph{two-dimensional} $\mathbb{Z}_2$ topological insulators. We further show that the low-energy bands near the band gaps can be perfectly captured by a modified Kane-Mele model Hamiltonian. Our work not only provides concrete two-dimensional materials that are very rare to realize \emph{two-dimensional} $\mathbb{Z}_2$ topological insulators, but also makes the graphene system to be realistic in hosting Kane-Mele type $\mathbb{Z}_2$ topological insulators.
\end{abstract}

\maketitle
\textit{Introduction---.} Two-dimensional (2D) $\mathbb{Z}_2$ topological insulators (TIs) are characterized by a pair of spin-helical counterpropagating edge modes, which are robust against non-magnetic disorder due to the protection of time-reversal symmetry~\cite{ref1,ref2}. This striking electronic transport phenomenon allows dissipationless electron transport without elastic backscattering, and thus offers a promising platform to realize novel low-power electronic and spintronic devices. 2D $\mathbb{Z}_2$ TI was first theoretically proposed by Kane and Mele in monolayer graphene~\cite{ref4}, and then realized in HgTe/CdTe as well as InAs/GaSb quantum wells~\cite {ref22,ref23,ref5}. Recent years have also witnessed abundant theoretically proposed candidates of 2D $\mathbb{Z}_2$ TIs~\cite {ref55,ref6,ref24,ref25,ref7,ref26,ref27,ref28,ref29,ref30,ref31,ref32,ref33,ref34,ref35,ref36,ref37,ref38,ref39}, among which only Bi-based thin films and Jacutingaite have been experimentally confirmed~\cite {ref44,ref45,ref46,ref47}. Nevertheless, observation of the 2D $\mathbb{Z}_2$ TI in graphene-based systems has not been experimentally realized yet. As the first and most appealing candidate for 2D $\mathbb{Z}_2$ TI, graphene has been attracting numerous focus for over fifteen years. However, although graphene exhibits various extraordinary properties~\cite {ref8}, its negligible spin-orbit coupling (SOC) precludes a sizeable band gap to realize the Kane-Mele type $\mathbb{Z}_2$ TI.

Several rewarding decorating approaches have been proposed to enhance the SOC in graphene, e.g., atomic adsorption~\cite {ref9, ref10, ref11}, atomic intercalation~\cite {ref40,ref41}, proximity effect~\cite {ref49}, or atomic embedding. All of them are challenging so far for experimental realization, except for embedding metal atoms, which are however mobile under irradiation to form a buckled lattice that breaks the flatness of graphene and thus loses its high mobility~\cite {ref59,ref60}. Recently, an improved strategy, i.e., embedding metal atoms together with nitrogen atoms into graphene, becomes practical, as some experiments have demonstrated~\cite {ref18,ref19,ref21} that the nitrogen atoms are able to anchor the metal atoms in the coplanar lattice of graphene, facilitating to form a stable 2D metal compound, named after $X$N$_4$-embedded graphene with $X$ being the metal atom. Compared with other types of modified graphene mentioned above, $X$N$_4$-embedded graphene has intriguing advantages that are demonstrated by both experiments and first-principles calculations, including that (i) it maintains a flat single-atom-thick structure as pristine graphene~\cite {ref18,ref19,ref21}; (ii) it covers twenty four kinds of metal elements with different atomic coverages~\cite{ref20}; (iii) its structural stability of CoN$_4$-embedded graphene has been ascertained via phonon spectra calculations and molecular dynamic simulations~\cite {ref12}. As the heavy transition metal atoms are enabled to be embedded into graphene, it may exhibit considerable SOC to realize the expectant topological states, e.g. $\mathbb{Z}_2$ TIs.

In this Letter, we construct a representative structural model of $X$N$_4$-embedded graphene and investigate its topological properties by using the density functional theory and tight-binding model calculations. We demonstrate that four kinds of 2D materials for $X$N$_4$-embedded graphene, as three of which (i.e., $X=$Pt, Ir, Rh) have been experimentally fabricated, are 2D $\mathbb{Z}_2$ TIs. We show that in the absence of SOC, they exhibit spin-degenerate gapless Dirac dispersions around two Dirac points that are away from K/K' points due to the atomic structural distortion. When SOC is further switched on, a sizeable band gap of 47/101 meV in (Pt/Os)N$_4$C$_{10}$ opens at the Fermi level, while a 46/25 meV gap in (Ir/Rh)N$_4$C$_{10}$ opens above the Fermi level. The $\mathbb{Z}_2$ index, edge states, spin Berry curvature and spin-Hall conductivity are calculated to confirm the topologically nontrivial properties. At the low-energy continuum limit, a modified Kane-Mele model is constructed to naturally captures the physical mechanism of the gapped Dirac cone in $X$N$_4$-embedded graphene. Finally, we illustrate that either the random distribution or the variable coverage of the embedding $X$N$_4$ units plays negligible role in the formation of Dirac cone structures, indicating that the observation of $\mathbb{Z}_2$ phase in $X$N$_4$-embedded graphene is experimentally feasible.

\begin{figure}[tb]
	\includegraphics[width=0.48\textwidth]{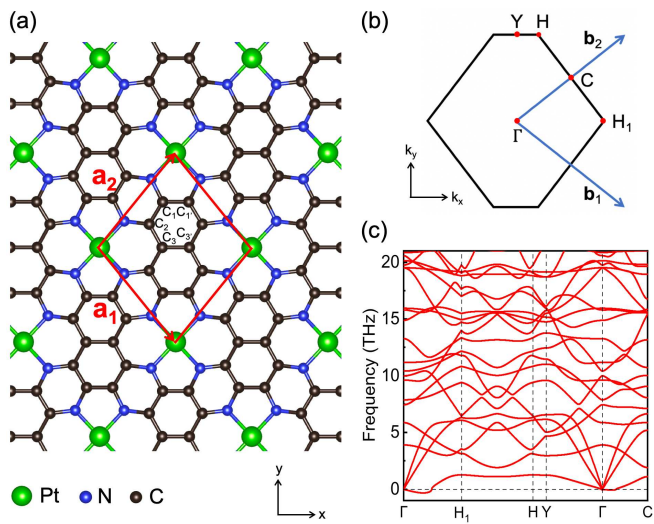}
	\caption{(a) Top view of the single layer ${\rm PtN_4C_{10}}$ lattice. The ${\rm {\bf a_1}}$ and ${\rm {\bf a_2}}$ are two basis vectors of the primitive cell. (b) The first Brillouin zone of the lattice and its high symmetry points. The ${\rm {\bf b_1}}$ and ${\rm {\bf b_2}}$ are two reciprocal basis vectors. (c) Phonon spectrum of the ${\rm PtN_4C_{10}}$ along high symmetry points.}
	\label{fig1}
\end{figure}

\begin{table}
	\centering
	\renewcommand\arraystretch{2}
	\begin{tabular}{ccccccccc}
		\hline \hline
		$d_{\rm C_{12}}$ & $d_{\rm C_{23}}$ & $d_{\rm C_{33'}}$ & $d_{\rm C_{11'}}$ & $d_{\rm C_1N}$ & $d_{\rm C_2N}$ & $d_{\rm NPt}$ \\
		\hline
		$1.389$ & $1.417$ & $1.456$ & $1.391$ & $1.363$ & $1.351$ & $1.915$ \\
		\hline \hline
	\end{tabular}
	\caption{Length of bonds between two atoms in ${\rm PtN_4C_{10}}$ in the unit of $\AA$.}
\label{tab-1}
\end{table}

\textit{Methods---.} Our first-principles calculations based on density functional theory are performed using the Vienna Ab initio Simulation Package (VASP) \cite {ref13, ref14} with the project augmented-wave method \cite{ref15} and exchange-correlation functional in the Perdew-Berke-Ernzerhofs form within the generalized-gradient approximation (GGA) \cite {ref16}. A vacuum buffer space over $16$ \AA is applied to minimize image interactions between adjacent slabs. All atoms are allowed to be fully relaxed until the force on each atom is less than $10^{-6}$ eV/\AA. The plane-wave energy cutoff is set to be 600 eV, with the first Brillouin-zone integration being carried out using a $15\times15\times1$ $\Gamma$-centered $k$-point grid. Phonon band dispersions are evaluated by using density functional perturbation theory based on the PHONOPY program \cite {ref42}. Maximally localized Wannier functions are constructed by using the WANNIER90 package \cite {ref17} to study their topological properties.

\textit{Structural and Electronic Properties---.} At each $X$N$_4$ unit, the central $X$ atom is connected with four N atoms. There are usually different coverage configurations of $X$N$_4$ units in graphene. Here, we choose a representative structure named after $X$N$_4$C$_{10}$, where $X$ is a transition metal atom. Hereinbelow, we take PtN$_4$C$_{10}$ (see Fig.~\ref{fig1}(a)) as a concrete example. In our calculations, all atoms in $X$N$_4$C$_{10}$ are coplanar to form a stable structure, indicating that it is a well-defined 2D system. Furthermore, the embedded PtN$_4$ units deform the pristine graphene and lead to the variation of the covalent bonds between carbon atoms, i.e., there are three inequivalent carbon atoms, ${\rm C_1}$, ${\rm C_2}$ and ${\rm C_3}$ in PtN$_4$C$_{10}$, with four different bonds as listed in Table~\ref{tab-1}. As a consequence, $X$N$_4$C$_{10}$ possesses the rhombic lattice with a plane symmetry group $Cmm$, affiliated to the point group $D_{2}$, which has the two-fold rotation symmetry along $x$/$y$/$z$ axis and is identical to the mirror symmetry of $z$-$x$/$y$-$z$ plane (denoted by $\mathcal{M}_{y/x}$) and the inversion symmetry, respectively.

\begin{figure*}
  \centering
  \includegraphics[width=\textwidth]{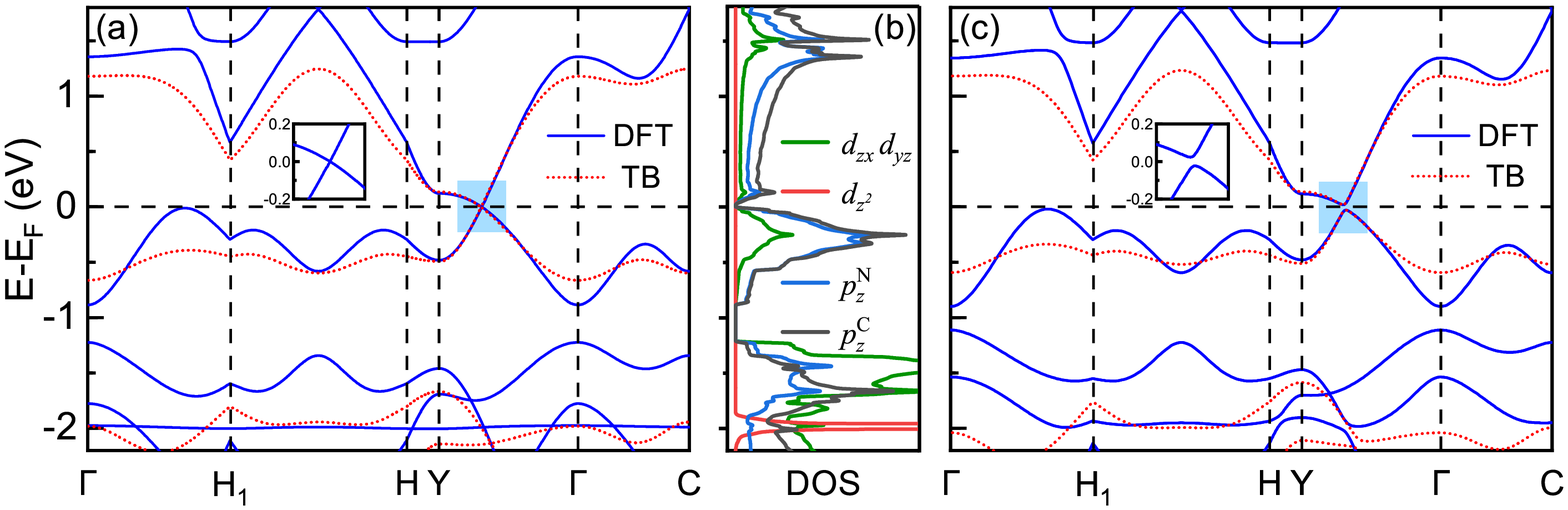}
  \caption{(a) and (c): Electronic structures of PtN$_4$C$_{10}$ obtained from first-principles calculations (blue solid curves) and tight-binding model (red dashed curves) (a) without SOC and (c) with SOC. (b) Projected density of states on the atomic orbitals. Insets of (a) and (c): Zooming in of band structures near Dirac points.}
	\label{fig2*}
\end{figure*}

Figure~\ref{fig1}(c) displays the phonon spectrum. One can see that there is no trace of imaginary frequency in the Brillouin zone, except a small portion of negative value near $\Gamma$ point. The limited negative imaginary frequency may arise from the difficulty of achieving numerical convergence for the flexural phonon branch, which usually appears in first-principles calculations for 2D materials~\cite {ref57,ref58}. Therefore, PtN$_4$C$_{10}$ can be considered to be dynamically stable.

The electronic structure of PtN$_4$C$_{10}$ shows a bulk band gap of tens of meV. In the absence of SOC, the valence and conduction bands touch at the Fermi level to form a distorted gapless Dirac cone, as displayed in Fig.~\ref{fig2*}(a). The density of states is vanishing at the Fermi level, as displayed in Fig.~\ref{fig2*}(b), which further confirms the existence of Dirac cone structures. In graphene, the two inequivalent Dirac cones at K/K' points are protected by the three fold rotational symmetry. A distortion of the honeycomb lattice can cause a shift of the Dirac cone in the (distorted) first Brillouin zone with lower spatial symmetry, which is commonly reported in deformed graphene, e.g., phagraphene~\cite{ref53}. For the electronic structure of PtN$_4$C$_{10}$, two Dirac points lie in the high symmetry lines between $\Gamma$ and Y, and $\Gamma$ and -Y, respectively. Therefore except along $k_x$ direction, the bands of PtN$_4$C$_{10}$ near the Dirac cone have anisotropic linear dispersions, meaning anisotropic Fermi velocities. Furthermore, the PtN$_4$ unit with $D_2$ symmetry generates a deformed square planar crystal field in the structural plane, which splits Pt $d$ orbitals into four distinct energy levels denoted by $E_{1g} (d_{zx}, d_{yz}), A_{1g} (d_{z^2}), B_{1g} (d_{x^2-y^2}), B_{2g} (d_{xy})$. The C $p_z$ and N $p_z$ orbitals form $\pi$ bonds together with Pt $d_{zx}$ and $d_{yz}$ orbitals, because they have the same odd parity, as their projected density of states are plotted in Fig.~\ref{fig2*}(b). The other $d$ orbitals with the even parity, on the other hand, do not couple with the above orbitals and thus form three discrete flat bands, in which two ($d_{x^2-y^2}$ and $d_{z^2}$) lie below the Fermi level and one ($d_{xy}$) lies above the Fermi level. After introducing SOC, a bulk band gap around 47 meV opens at the Dirac point as displayed in Fig.~\ref{fig2*}(c), indicating an insulating state.

\begin{figure}[tb]
	\includegraphics[width=0.48\textwidth]{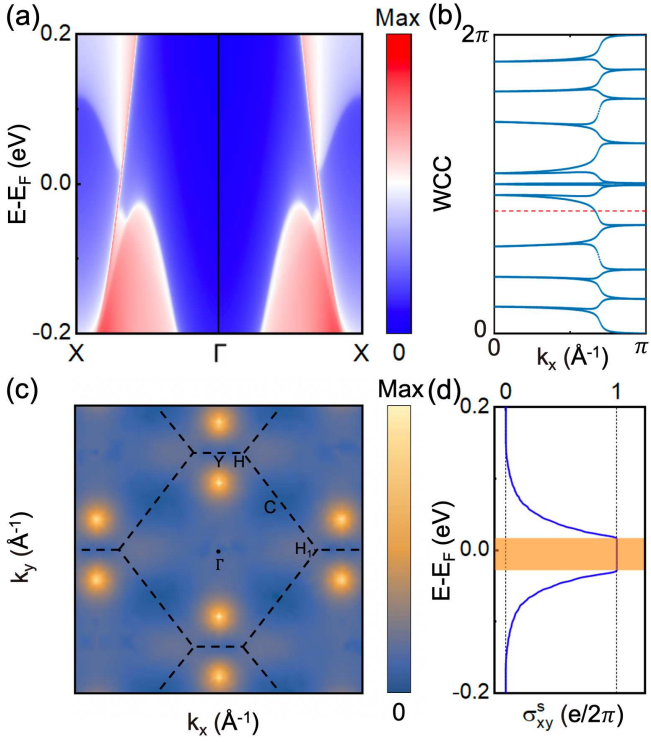}
	\caption{(a) Edge densities of states for ${\rm PtN_4C_{10}}$. (b) Wannier center charge (WCC) evolution (Wilson loop) curves. (c) Distribution of spin Berry curvature in the Brillouin zone. (d) Spin-Hall conductivity. (a) and (b) are obtained from first-principles calculations, while (c) and (d) are calculated from the tight-binding model with the parameters listed in Table II.}
	\label{fig3}
\end{figure}

\textit{Topological Properties---.} The gapped PtN$_4$C$_{10}$ is a 2D $\mathbb{Z}_2$ TI. The $\mathbb{Z}_2$ topological number can be evaluated from the Wannier charge center (WCC)~\cite {ref50,ref51}. The evolution curves of WCC are obtained by using the approach implemented in WannierTools packages~\cite {ref43} [see Fig.~\ref{fig3}(b)], where the curves cross the reference line odd times, indicating $\mathbb{Z}_2=1$. We further calculate the edge densities of states by means of Green's function technique in a semi-finite ribbon with armchair edges, as displayed in Fig.~\ref{fig3}(a). The helical edge states connect the bulk conduction and valence bands, which characterizes the 2D $\mathbb{Z}_2$ TIs.

\begin{table}
	\centering
	\renewcommand\arraystretch{2}
	\begin{tabular}{ccccccccc}
		\hline \hline
		$\epsilon^C_{p_{z}}$ & $\epsilon^N_{p_{z}}$ & $\epsilon^{Pt}_{d_{xz}}$ & $\epsilon^{Pt}_{d_{yz}}$ & $V^{CC}_{pp\pi}$ & $V^{CN}_{pp\pi}$ & $V^{NPt}_{pd\pi}$ & $t_{\rm SO}$ \\
		\hline
		$0.15$ & $-1.17$ & $2.35$ & $6.52$ & $-5.60$ & $2.70$ & $1.15$ & $1.53$ \\
		\hline \hline
	\end{tabular}
	\caption{Slater-Koster tight-binding parameters of ${\rm PtN_4C_{10}}$ in the presence of SOC. Unit: eV.}
    \label{tab-2}
\end{table}

\textit{Underlying Physics from Tight-binding Model---.} To further understand the physical mechanism of the $\mathbb{Z}_2$ phase of PtN$_4$C$_{10}$, we construct a tight-binding model by using Slater-Koster approximation on the orthogonal basis of \{$|p^{\rm C}_{z}\rangle$, $|p^{\rm N}_{z}\rangle$, $|d_{zx}\rangle, |d_{yz}\rangle$\} $\otimes$ \{$|\uparrow\rangle$\, $|\downarrow\rangle$\}. The tight-binding model Hamiltonian can be written as:
\begin{eqnarray}
\begin{aligned}
  H_{\rm TB}=&\sum_{i\alpha}\epsilon_{\alpha}c_{i\alpha}^{\dagger}c_{i\alpha}-\sum_{\langle i,j \rangle}t_{ij,\alpha\beta}c_{i\alpha}^{\dagger}c_{j\beta}\\
  	&+t_{\rm SO}\sum_{i\alpha} c_{i\alpha}^{\dagger}\bm{l}\cdot\bm{s}c_{i\beta},
\end{aligned}
\end{eqnarray}
where $c^\dagger_{i\alpha}=(c^\dagger_{i\alpha\uparrow},c^\dagger_{i\alpha\downarrow})$ denotes the creation operator of an electron at the $i$-th atomic site with $\uparrow$/$\downarrow$ and $\alpha$/$\beta$ representing spin up/down and different orbitals, respectively. The first term stands for the on-site energy. The second term is the nearest-neighbor hopping with an amplitude of $t_{ij,\alpha\beta}$, which are functions of the Slater-Koster parameters listed in Table~\ref{tab-2}. The third term represents the atomic SOC of Pt $d$ orbitals with a strength of $t_{\rm SO}$. $\bm{s}=(s_x, s_y, s_z)$ and $\bm{l}=(l_x, l_y, l_z)$ are the Pauli matrices and orbital-angular-momentum operators, respectively. Due to the lattice deformation, the hopping amplitude between carbon atoms changes with the interatomic distance and can be determined by $t_{ij}=V^{CC}_{pp\pi}e^{-\beta(d_{ij}/d_0-1)}$, where $V^{CC}_{pp\pi}$ is the Slater-Koster parameter and $d_{ij}$ is the distance between two carbon atoms. Here, we set $\beta=2.20$ and $d_0=1.42$ \AA. Using these fitting parameters that are extracted from first-principles calculations, one can obtain the band structure that agrees quantitatively with the first-principles results near the Dirac point, as shown in Figure~\ref{fig2*}. When $t_{\rm SO}=0$, a gapless Dirac cone band is formed; when $t_{\rm SO}\neq 0$, a band gap opens at the Dirac point.

To further verify the topologically nontrivial band gap, we first calculate the spin Berry curvature $\Omega^s(\textbf{k})$, which is well-defined for flattened single-atom-thick systems with the mirror symmetry $\mathcal{M}_z$ about the $x$-$y$ plane:
\begin{eqnarray}
\begin{aligned}
  &\Omega^s(\textbf{k})=\sum_n f_n\Omega^s_n(\textbf{k}),\\
  &\Omega^s_n(\textbf{k})=-2{\rm Im}\sum_{m\neq n}\frac{\langle{\psi_{n\textbf{k}}}|j_x|\psi_{m\textbf{k}}\rangle{\langle{\psi_{m\textbf{k}}}|v_y|\psi_{n\textbf{k}}\rangle}}{(\varepsilon_m-\varepsilon_n)^2},
  \end{aligned}
\end{eqnarray}
where $\psi_{n\textbf{k}}$ and $\varepsilon_n$ are respectively the Bloch eigenstate and eigenvalue of the $n$-th band at $\textbf{k}$, $f_n$ is the Fermi-Dirac distribution function, and $j_x$ is the spin current operator defined as $\frac12(s_zv_x+v_xs_z)$ with $v_{x(y)}$ being the velocity operator and $s_z$ the spin operator. Figure~\ref{fig3}(c) plots the spin Berry curvature of all the occupied bands. The sharp peaks appear at the positions of Dirac gap in the Brillouin zone. The spin-Hall conductivity directly related to the spin Berry curvature can be thus evaluated via the Kubo formula~\cite {ref48}:
\begin{eqnarray}
\sigma_{xy}^s=\frac{e}{(2\pi)^2}\int_{\rm BZ}\Omega^s(\textbf{k})d^2\textbf{k}.
\end{eqnarray}
Figure~\ref{fig3}(d) plots the spin-Hall conductivity $\sigma_{xy}^s$ as a function of Fermi energy ($E_{\rm F}$) for PtN$_4$C$_{10}$. One can observe that $\sigma_{xy}^s$ keeps quantized whenever $E_{\rm F}$ lies inside the band gap.

\textit{Low-Energy Effective Model---.} In the low-energy limit, the electronic structure can be captured by the Kane-Mele model constraint by the symmetry of the system, which is even under mirror, parity and time reversal~\cite{ref3,ref4,ref55,ref56}. Nevertheless, due to the decrease of the spatial symmetry, the fermi velocity is no longer isotropic. We thus modify the Kane-Mele
model as $H_{\rm eff}=H_0+H_{\rm SO}$ around the Dirac points with $H_0=Ak_x\sigma_x\tau_z+Bk_y\sigma_y+Ck_y\tau_z$ and $H_{\rm SO}=\Delta_{\rm SO}\sigma_z\tau_zs_z$. Here, $\vec\sigma$ denotes the Pauli matrices, $\tau_z=\pm1$ describes the states at two Dirac points in the first Brillouin zone [Fig.~\ref{fig3}(c)], and $s_z$ represents electron's spin. A, B and C are parameters in the unit of eV. The pair of gapless Dirac cones are protected by $\mathcal{M}_x$ and $\mathcal{M}_y$, which thus allow an additional term that is proportional to $k_y\tau_z$. $H_{\rm SO}$ is the intrinsic SOC in the Kane-Mele model, which generates a topologically nontrivial indirect gap $2\sqrt{1-C^2/B^2} \Delta_{\rm SO}$, as long as $C^2<B^2$. By fitting parameters, the bands exhibit good agreement with first-principles calculations (see Supplemental Materials~\cite {ref52}). Since our calculations of PtN$_4$-embedded graphene with different coverages suggest that the Dirac cone retains with the decrease of coverages~\cite {ref52}, the effective model offers a very precise description of the Dirac cone in the low-energy range, and captures its odd $\mathbb{Z}_2$ phase. Thus, we show that PtN$_4$-embedded graphene belongs to the modified Kane-Mele type 2D $\mathbb{Z}_2$ TI.

\begin{table}
	\centering
	\begin{tabular}{cccccc}
	\hline \hline
	\begin{tabular}[c]{@{}l@{}} Metal\\ \makecell[c]{atom}\end{tabular}
		 & \begin{tabular}[c]{@{}l@{}} \makecell[c]{$a$ (\AA)} \end{tabular} & \begin{tabular}[c]{@{}l@{}} \makecell[c]{$\theta$} \end{tabular}& \begin{tabular}[c]{@{}l@{}} \makecell[c]{$\Delta$(meV)} \end{tabular} & \begin{tabular}[c]{@{}l@{}} \makecell[c]{Synthesized?~\cite{ref20}} \end{tabular}\\
		\hline
		\makecell[c]{${\rm Ni}$} & \makecell[c]{$6.476$} &$102.33^{\circ}$ & \makecell[c]{$0.0$} & \makecell[c]{Y}\\
		\makecell[c]{${\rm Zn}$} & \makecell[c]{$6.590$} &$102.52^{\circ}$ & \makecell[c]{$0.0$} & \makecell[c]{N}\\
		\makecell[c]{${\rm Rh}$} & \makecell[c]{$6.595$} &$103.09^{\circ}$& \makecell[c]{$24.8$} & \makecell[c]{Y}\\
		\makecell[c]{${\rm Pd}$} & \makecell[c]{$6.591$} &$102.92^{\circ}$& \makecell[c]{$0.0$}& \makecell[c]{Y}\\
		\makecell[c]{${\rm Ag}$} & \makecell[c]{$6.639$} &$102.79^{\circ}$& \makecell[c]{$0.0$} & \makecell[c]{Y}\\
		\makecell[c]{${\rm Cd}$} & \makecell[c]{$6.727$} &$103.06^{\circ}$& \makecell[c]{$0.0$} & \makecell[c]{Y}\\
		\makecell[c]{${\rm Os}$} & \makecell[c]{$6.614$} &$103.24^{\circ}$& \makecell[c]{$100.9$} & \makecell[c]{N}\\
		\makecell[c]{${\rm Ir}$} & \makecell[c]{$6.603$} &$103.28^{\circ}$& \makecell[c]{$46.3$} & \makecell[c]{Y}\\
		\makecell[c]{${\rm Pt}$} & \makecell[c]{$6.600$} &$103.18^{\circ}$&  \makecell[c]{$47.1$} & \makecell[c]{Y}\\
		\makecell[c]{${\rm Au}$} & \makecell[c]{$6.613$} &$102.08^{\circ}$& \makecell[c]{$0.0$} & \makecell[c]{Y}\\
		\hline \hline
	\end{tabular}
	\caption{Properties of ${X\rm N_4C_{10}}$ with $X$ being the metal atoms. $a$ and $\theta$ are the lattice constants, $\Delta$ is the global bulk band gap. Above structures are all non-magnetic and single-atom-thick.}
\label{tab-3}
\end{table}

\textit{Other Materials Candidates---.} In Table~\ref{tab-3}, we list the parameters of dozen systems of $X$N$_4$C$_{10}$, with $X$ being the transition metal atoms. One can find that $X$N$_4$C$_{10}$ ($X$= Ir, Rh, and Os) with the same rhombic crystal structure also have sizeable nontrivial bulk gaps. The corresponding structural stability, electronic properties and topological properties are further provided in Supplemental Materials~\cite{ref52}. It is worth mentioning that the structure of OsN$_4$C$_{10}$ is mechanically stable via our phonon spectrum calculation, and has a large topologically nontrivial gap around 101 meV at the Fermi level, though the experimental synthesis of OsN$_4$-embedded graphene has not been reported yet~\cite{ref20}. Furthermore, other $X$N$_4$C$_{10}$ ($X$= Ni, Zn, Pd, Ag, Cd, and Au) also host the locally gapped Dirac cone structures, which are however overwhelmed by the first valance bands that cross the Fermi level. Such $\mathbb{Z}_2$ metallic states, however, could be tuned into the normal $\mathbb{Z}_2$ insulating states via doping non-magnetic disorder.

\textit{Random embedding distribution---.} In realistic synthesized compounds in experiments~\cite {ref18,ref19,ref21,ref20}, graphene is no longer subjected to periodically distributed embedding $X$N$_4$ units. It is thus reasonalbe to explore the effect of random distribution on their topological properties. From above analysis, the $X$N$_4$ units here play an essential role as enhancing intrinsic SOC, which is very similar to that of transition metal atoms adsorbed on graphene~\cite{ref11,ref54}. In that case, random adatom distribution helps weaken inter-valley scattering while it has a negligible effect on the SOC. Besides, in our calculations~\cite{ref52}, the Dirac cone structures are still retained at low coverages of the $X$N$_4$ units, hence the odd $\mathbb{Z}_2$ phase is expected to be stable in the presence of randomly distributed embedding $X$N$_4$ units.

\textit{Summary---.} We systematically investigate the topological properties of $X{\rm N_4}$-embedded graphene. We demonstrate that ${X\rm N_4C_{10}}$ ($X$= Pt, Ir, Rh, Os) are 2D $\mathbb{Z}_2$ TIs. The embedded transition metal maintains the mechanic stability of graphene and modifies the Dirac cone structures of graphene to foster the stronger SOC. Without SOC, there are shifted Dirac bands with a vanishing band gap near the Fermi level. After introducing SOC, a topologically nontrivial band gap opens, characterized by the odd $\mathbb{Z}_2$ and topological edge states. A tight-binding model is constructed to reveal that the atomic SOC can induce such a nontrivial band gap. The low-energy bands of the gapped Dirac cone are described by a modified Kane-Mele model with the lower symmetry due to the deformation of graphene, indicating $X{\rm N_4}$-embedded graphene is 2D Kane-Mele type TIs. The bulk gap retains with a lower coverage of metal atoms, which is experimentally feasible. Our study of this system opens up routes towards exploring more graphene-based 2D $\mathbb{Z}_2$ TIs.

\begin{acknowledgments}
We thank Y. Ren and Y. Han for helpful discussions. This work was financially supported by the National Key Research and Development Program (2017YFB0405703), the National Natural Science Foundation of China (11974327), Anhui Initiative in Quantum Information Technologies, and the Fundamental Research Funds for the Central Universities. We are grateful to AMHPC and Supercomputing Center of USTC for providing the high-performance computing resources.
\end{acknowledgments}

\end{document}